\documentclass[preprint,preprintnumbers,amsmath,amssymb,nofootinbib]{revtex4}
\usepackage{booktabs}
\usepackage{mathrsfs}
\usepackage{epsfig}
\usepackage{graphicx}
\usepackage{dcolumn}
\usepackage{bm}
\usepackage{amsmath}
\usepackage{multirow}
\usepackage{subfigure}
\usepackage{color}
\usepackage{slashed}

\let\jnfont=\rm
\def\NPB#1,{{\jnfont Nucl.\ Phys.\ B }{\bf #1},}
\def\PLB#1,{{\jnfont Phys.\ Lett.\ B }{\bf #1},}
\def\EPJC#1,{{\jnfont Eur.\ Phys.\ Jour.\ C }{\bf #1},}
\def\PRD#1,{{\jnfont Phys.\ Rev.\ D }{\bf #1},}
\def\PRL#1,{{\jnfont Phys.\ Rev.\ Lett.\ }{\bf #1},}
\def\MPLA#1,{{\jnfont Mod.\ Phys.\ Lett.\ A }{\bf #1},}
\def\JPG#1,{{\jnfont J.\ Phys.\ G}{\bf #1},}
\def\CTP#1,{{\jnfont Commun.\ Theor.\ Phys.\ }{\bf #1},}
\def\ZPC#1,{{\jnfont Z.\ Phys.\ C }{\bf #1},}
\def\JHEP#1,{{\jnfont JHEP \ }{\bf #1},}
\def\Rv{\not{\hbox{\kern-1pt $R$}}}
\def\p{\not{\hbox{\kern-3pt $p$}}}

\newcommand{\bea}{\begin{eqnarray}}
\newcommand{\eea}{\end{eqnarray}}

\newcommand{\bcen}{\begin{center}}
\newcommand{\ecen}{\end{center}}

\newcommand{\beq}{\begin{eqnarray}}
\newcommand{\eeq}{\end{eqnarray}}

\newcommand{\tabincell}[2]{\begin{tabular}{@{}#1@{}}#2\end{tabular}}

\def\simlt{\mathrel{\raise.3ex\hbox{$<$\kern-.75em\lower1ex\hbox{$\sim$}}}}
\def\simgt{\mathrel{\raise.3ex\hbox{$>$\kern-.75em\lower1ex\hbox{$\sim$}}}}

\def\t1{\tilde{t_1}}

\begin{document}

\preprint{
\begin{minipage}[b]{0.75\linewidth}
\begin{flushright}
 \end{flushright}
\end{minipage}
}

\title{Probing stops in the coannihilation region at the HL-LHC: \\
a comparative study of different processes}
\author{ Guang Hua Duan$^{1,2}$}
\author{ Xiang Fan$^{1,2}$}
\author{ Ken-ichi Hikasa$^{3}$}
\author{ Bo Peng$^{1,2}$}
\author{ Jin Min Yang$^{1,2,3}$}

\affiliation{
 $^1$ CAS Key Laboratory of Theoretical Physics, Institute of Theoretical Physics,
Chinese Academy of Sciences, Beijing 100190, China \\
 $^2$ School of Physical Sciences, University of Chinese Academy of Sciences, Beijing 100049, China \\
 $^3$ Department of Physics, Tohoku University, Sendai 980-8578, Japan
 \vspace*{1.5cm}
 }%


\begin{abstract}
In the minimal supersymmetric model, the coannihilation of the lighter stop
$\tilde{t}_1$ and bino-like dark matter $\chi$ provides a feasible way
to accommodate the correct dark matter relic abundance.
In this scenario, due to the compressed masses, $\tilde{t}_1$ merely appears
as missing energy at the LHC and thus the pair production of $\tilde{t}_1$
can only be probed by requiring an associated energetic jet.
Meanwhile, since $\tilde{t}_2$ and $\tilde{b}_1$ are correlated in mass and
mixing with $\tilde{t}_1$, the production of $\tilde{t}_2\tilde{t}_2^*$ or
$\tilde{b}_1\tilde{b}_1^*$, each of which dominantly decays into
$\tilde{t}_1$ plus $Z$, $h$ or $W$ boson, may serve as a complementary probe.
We examine all these processes at the HL-LHC and find that the $2\sigma$
sensitivity to $\chi$ mass can be as large as about 570 GeV, 600 GeV and 1.1 TeV
from the production process of
$\tilde{t}_1\tilde{t}_1^*+{\rm jet}$, $\tilde{t}_2\tilde{t}_2^*$ and
$\tilde{b}_1\tilde{b}_1^*$, respectively.

\end{abstract}

\maketitle

\section{Introduction}
\label{sec1}
The nature of dark matter (DM) remains a mystery in particle physics.
In  minimal supersymmetric standard model (MSSM) with conserved $R-$parity,
the lightest neutralino $\chi$  can serve as a DM candidate. However, the null results
of DM direct detections \cite{pandaX,xenon-1t,Akerib:2017kat} give significant constraints
on the neutralino sector in the MSSM.  It is notable that the stop-bino coannihilation,
in which DM is the bino-like lightest supersymmetric particle (bino-LSP) and the
stop ($\tilde{t}_1$) is the next-to-lightest supersymmetric particle (NLSP)
and nearly degenerate with the bino-LSP,  provides a feasible mechanism to accommodate
the DM relic abundance. Because of the extremely weak interaction between the bino-LSP
and nucleons, this scenario can easily evade the DM direct detection
constraints \cite{Pierce:2017suq}.
However, the search of stops at the LHC in this scenario is rather
challenging\footnote{The search of stops at the LHC has been a hot topic and
numerous studies have been performed in various cases, e.g.,
the large or small stop-top or stop-LSP mass splitting \cite{stop-lhc1,stop-lhc2,stop-lhc3},
the single stop production \cite{stop-lhc4},
the stop in natural SUSY \cite{stop-lhc5},
machine learning in stop production  \cite{stop-lhc6} and
other miscellaneous cases \cite{stop-lhc7}.}.
The reason is that due to the compressed masses,  $\tilde{t}_1$ is merely appearing
as missing energy and the pair production of $\tilde{t}_1$ can only be probed by requiring an
associated energetic jet.

On the other hand, we should note that $\tilde{t}_2$ and $\tilde{b}_1$ are correlated
with $\tilde{t}_1$
since $\tilde{t}_{L,R}$ mix into mass eigenstates $\tilde{t}_{1,2}$ (see the following section)
while $\tilde{b}_L$ ($\tilde{b}_1=\tilde{b}_L$, neglecting the sbottom mixing)
has the same soft mass as $\tilde{t}_L$.
Furthermore, to avoid fine-tuning, these particles should not be too
heavy\footnote{Note that the stops cannot be too light in order to
give the 125 GeV Higgs mass except a singlet is introduced \cite{higgs-susy}.}
because at one-loop level we approximately have \cite{Papucci:2011wy,Baer:2016bwh}
\begin{eqnarray}
\Delta\equiv\frac{\delta m_h^2}{m_h^2}=\frac{3y_t^2}{4\pi^2m_h^2}(m_{Q_3}^2+m_{U_3}^2
+A_t^2)\log\frac{\Lambda}{m_{\rm SUSY}}
\label{naturalness}
\end{eqnarray}
where $m_{\rm SUSY}=\sqrt{m_{\tilde{t}_1}m_{\tilde{t}_2}}$, $\Lambda$ is the cut-off scale,
$Q_3=(\tilde{t}_L,\tilde{b}_L)$ and $U_3=\tilde{t}_R$.
Therefore, the production of
$\tilde{t}_2\tilde{t}_2^*$ or $\tilde{b}_1\tilde{b}_1^*$, followed by the dominant decays into
$\tilde{t}_1$ plus $Z$, $h$ or $W$ boson, may serve as a complementary probe of stops in
such a stop-bino coannihilation scenario.

In this work we perform a comprehensive study for all these correlated processes
at the HL-LHC (14 TeV, 3000 fb$^{-1}$). We will first perform a scan to figure out the
 stop-bino coannihilation parameter space. Then we display the properties of
$\tilde{t}_{1,2}$ and  $\tilde{b}_1$ in this stop-bino coannihilation parameter space.
For the $\tilde{t}_1\tilde{t}_1^*+{\rm~jet}$ production which has been searched at the LHC,
we will show its current sensitivity and then extend the coverage to the HL-LHC.
For the productions $\tilde{t}_2\tilde{t}_2^*$ and $\tilde{b}_1\tilde{b}_1^*$,
followed the dominant decays $\tilde{t}_2 \rightarrow \tilde{t}_1 + Z/h$
and $\tilde{b}_1 \rightarrow \tilde{t}_1 + W$,
we will examine the HL-LHC sensitivities through Monte Carlo simulations
of the signals and backgrounds.

The structure of this paper is organized as follows.
In Sec.~II, we briefly review stop-bino coannihilation scenario
and discuss the details of our scan. In Sec.~III, we perform detailed Monte Carlo simulations
for the productions of  $\tilde{t}_1\tilde{t}_1^*+{\rm jet}$, $\tilde{t}_2\tilde{t}_2^*$
and $\tilde{b}_1\tilde{b}_1^*$ at the HL-LHC. Finally, we give our conclusions in Sec.~IV.

\section{Stop-bino coannihilation}
In the MSSM, the mass matrix of stop sector in gauge-eigenstate basis ($\tilde{t}_L$,$\tilde{t}_R$)
is given by
\begin{eqnarray}
M_{\tilde{t}}^2=
\left(
\begin{array}{cc}
m_{\tilde{t}_L}^2 &m_tX_t^{\dag}\\
m_tX_t& m_{\tilde{t}_R}^2\\
\end{array}
\right)
\end{eqnarray}
where
\begin{eqnarray}
&&m_{\tilde{t}_L}^2=m_{\tilde{Q}_{3L}}^2+m_t^2+m_Z^2\left(\frac{1}{2}
-\frac{2}{3}\sin^2\theta_W\right)\cos2\beta,\\
&& m_{\tilde{t}_R}^2=m_{\tilde{U}_{3R}}^2+m_t^2+\frac{2}{3}m_Z^2\sin^2\theta_W\cos2\beta,\\
&& X_t = A_t -\mu \cot\beta.
\end{eqnarray}
The mixing between $\tilde{t}_L$ and $\tilde{t}_R$ is induced by $X_t=A_t-\mu\cot\beta$,
where $A_t$ is the stop soft-breaking trilinear coupling.  One can diagonalize the mass
matrix through a rotation
\begin{eqnarray}
\left(
\begin{array}{c}
\tilde{t}_1 \\
\tilde{t}_2 \\
\end{array}
\right)
=
\left(
\begin{array}{cc}
\cos\theta_{\tilde{t}} &\sin\theta_{\tilde{t}}\\
-\sin\theta_{\tilde{t}}& \cos\theta_{\tilde{t}}\\
\end{array}
\right)
\left(
\begin{array}{c}
\tilde{t}_L \\
\tilde{t}_R \\
\end{array}
\right),
\end{eqnarray}
where $\tilde{t}_1$ and $\tilde{t}_2$ are the mass eigenstates of lighter and heavier stops,
respectively.
The mixing angle $\theta_{\tilde{t}}$ between $\tilde{t}_L$ and $\tilde{t}_R$ is determined by
\begin{eqnarray}
{\tan 2 \theta_{\tilde{t}}}=\frac{2m_tX_t}{m_{\tilde{t}_L}^2-m_{\tilde{t}_R}^2}.
\end{eqnarray}

In the early universe,
the freeze-out number density for the bino-LSP DM will be over-abundant because the annihilation
cross section $\sigma$ in the Boltzmann equation is too small to keep DM thermal equilibrium with
SM particles for sufficient time\footnote{The $Z/h$ funnel as another exception is that the annihilation
cross section could be enhanced when bino-LSP mass becomes half of $m_{Z/h}$.}.
When the stop ($\tilde{t}_1$) mass is close to bino-LSP mass, the annihilation
cross section $\sigma$ is replaced by the effective cross section \cite{Griest:1990kh}
 \begin{eqnarray}
\sigma_{\rm eff}=\sum_{ij}\sigma_{ij}  r_ir_j
\end{eqnarray}
with
 \begin{eqnarray}
r_i=\frac{g_i(1+\Delta_i)^{3/2}e^{-\Delta_i/T}}{\sum\limits_k g_k(1+\Delta_k)^{3/2}e^{-\Delta_k/T}}
\end{eqnarray}
where $\Delta_i=(m_i-m_\chi)/T$, $m_i$ and $g_i$ are the mass and degrees of freedom of the
particle $i=\{\chi,\tilde{t}_1\}$, and $\sigma_{ij}$ denotes the cross section of particle $i$
annihilating with particle $j$. The annihilation modes of $\tilde{t}_1$  with $\chi $ or
itself can enhance
$\sigma_{\rm eff}$ if $\tilde{t}_1$ is nearly degenerate with the bino-LSP.
We can also see that $\sigma_{\tilde{t}_1\tilde{t}_1}$ is  suppressed by double exponents compared
to $\sigma_{\chi\chi}$, while $\sigma_{\tilde{t}_1\chi}$ is suppressed by single exponent.
Therefore, when the mass splitting $\Delta_{\tilde{t}_1}$ is small, the contribution to relic abundance
from the $\tilde{t}_1\chi$ annihilation tends to be more important than that from the
$\tilde{t}_1\tilde{t}^\ast_1$ annihilation, although this also depends on their respective
cross section.

In order to obtain the stop-bino coannihilation parameter space, we
use \textsf{SuSpect 2.41} \cite{Djouadi:2002ze} to calcualte the mass spectrum
and \textsf{SDECAY 1.5} \cite{Muhlleitner:2003vg} to evaluate sparticle decay width and branching ratio.
We regard the lighter stop as right-handed dominated. The reason for such assumption
is that if $m_{\tilde{t}_R}=m_{\tilde{t}_{L}}$ at some high energy scale, $m_{\tilde{t}_R}$ tends to be
smaller than $m_{\tilde{t}_L}$ at the electroweak scale from the renormalization group equations (RGE)
evolution \cite{Delgado:2012eu}. The stop mixing angle  $\cos^2\theta_{\tilde{t}}\simlt0.5$ is required
so that the lighter stop $\tilde{t}_1$ is right-handed dominant. Except for $\{M_1,m_{Q_3},m_{U_3},A_t\}$,
other soft-breaking masses (including the CP-odd Higgs mass $M_A$) and trilinear couplings are
set to 5 TeV and zero, respectively. The higgsino mass parameter $\mu$ and $\tan{\beta}$ are chosen
as 3 TeV and 20. The \textsf{micrOMEGAs 4.3.5} \cite{Barducci:2016pcb} is used to compute the DM
relic abundance $\Omega_{\chi}h^2$.

In our scan we impose the following constraints\footnote{Here we do not require SUSY to
explain the muon g-2 anomaly, which requires light sleptons \cite{susy-gm2}}:
\begin{itemize}
	\item[(i)] The lighter CP-even Higgs mass is required to be in the range of
$125\pm3$ GeV \cite{higgs-atlas,higgs-cms}.
	\item[(ii)]The DM relic abundance satisfies the observed
value $\Omega_{\chi}h^2=0.1186\pm0.0020$ within 2$\sigma$ range \cite{relic density}.
	\item[(iii)] To avoid the existence of a color or charge breaking vaccum deeper than
the electroweak vacuum in the scalar potential, the trilinear coupling $A_t$ should not exceed
the upper bound $A_t^2 \simlt2.67(m_{\tilde{t}_L}^2+m_{\tilde{t}_R}^2+\mu^2+m_{H_u}^2)$ \cite{Chowdhury:2013dka}.
\end{itemize}

\begin{figure}[ht!]
	\centering
	\includegraphics[height=3.3in,width=3.3in]{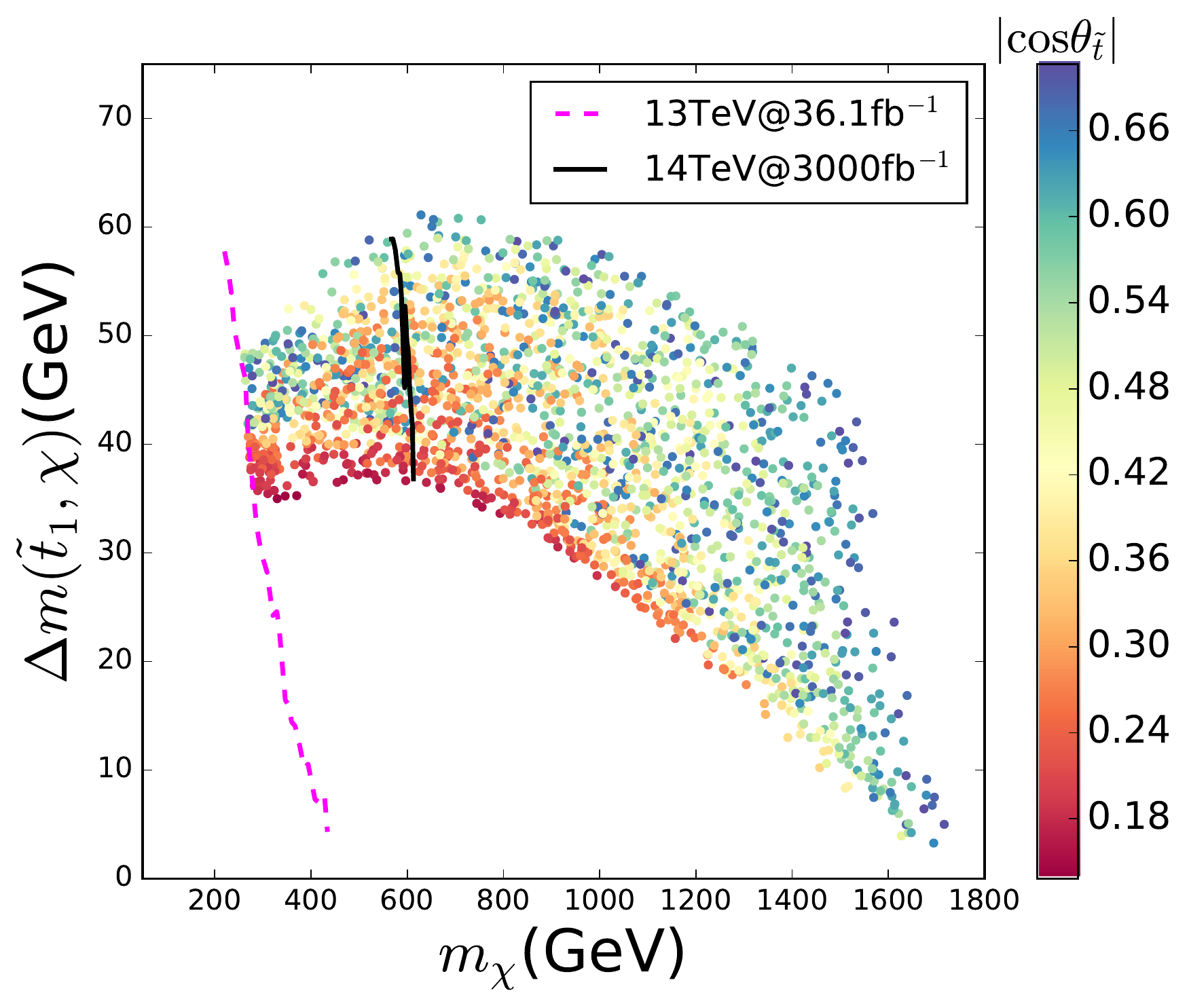}
	\includegraphics[height=3.3in,width=3.0in]{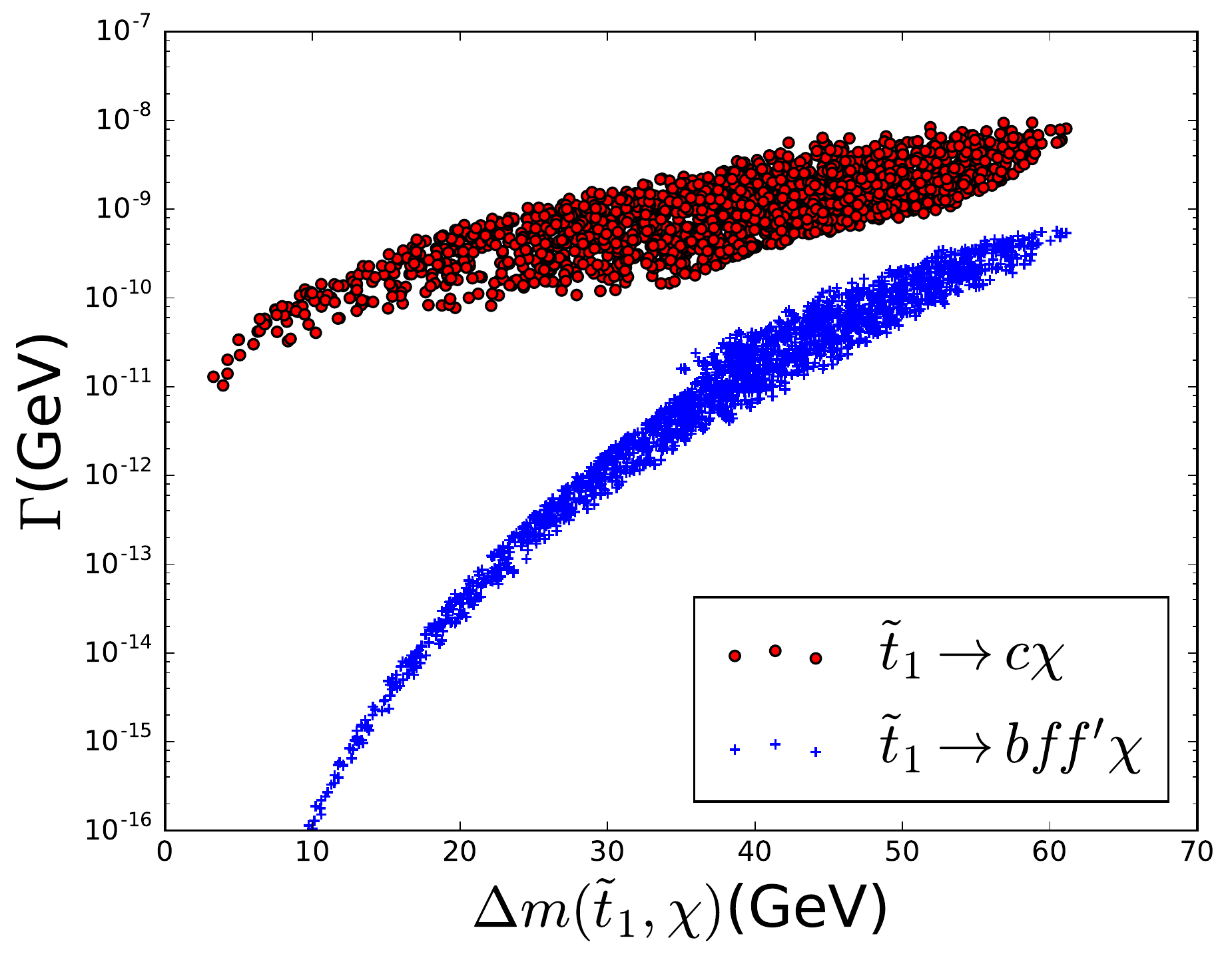}
	\vspace{-0.5cm}
	\caption{Scatter plots of the stop-bino coannihilation parameter space satisfying the
constraints (i)--(iii).
The left panel shows the DM mass $m_\chi$ versus mass splitting $\Delta m(\tilde{t}_1,\chi)$ with
the colormap denoting the size of  $|\cos\theta_{\tilde{t}}|$.
The dashed magenta curve and the solid black curve are the $2\sigma$ sensitivities of
the $\tilde{t}_1 \tilde{t}_1^*+{\rm jet}$ production from the current ATLAS search  \cite{Aaboud:2017phn}
and our simulations for the HL-LHC, respectively.
The right panel shows the stop $\tilde{t}_1$ decay width of the four-body channel
$\tilde{t}_1\rightarrow bf \bar f^\prime \chi$  and the FCNC two-body channel  $\tilde{t}_1\rightarrow c \chi$.
}
	\label{fig1}
\end{figure}

In the left panel of Fig.~\ref{fig1}, we display the stop-bino coannihilation parameter space
that satisfies the constraints (i)--(iii), where the $B$ physics constraints are ignored because
of the decoupled higgsino mass parameter, and the contribution of the stops to
$h\rightarrow\gamma\gamma$ (and $gg$) \cite{Delgado:2012eu} is also negligibly small.
We can see that the mass splitting $\Delta m(\tilde{t}_1,\chi)$ increases with $|\cos\theta_{\tilde{t}}|$
because the component of left-handed stop annihilates with itself more efficiently due
to the SU(2)$_L$ interaction. Besides, it can be seen that the maximal value of $m_{\chi}$ is
about $1.8$ TeV, where $\tilde{t}_1\tilde{t}^*_1\rightarrow gg$ is the dominant annihilation
mode because of the QCD interaction and small mass splitting $\Delta m(\tilde{t}_1,\chi)$
or small $\Delta_{\tilde{t}_1}$.

 \section{Probing stops in the coannihilation region at the HL-LHC}

\subsection{The $\tilde{t}_1 \tilde{t}^{\ast}_1+ {\rm jet}$ production}
Since the lighter stop is nearly degenerate with the bino-LSP, the two-body decay
channel $\tilde{t}_1\rightarrow t \chi $ and three-body decay channel $\tilde{t}_1\rightarrow  bW\chi$
are kinematically forbidden. The lighter stop will dominantly decay via the four-body channel
$\tilde{t}_1\rightarrow  bf\bar f' \chi$  and loop induced flavor-changing neutral current (FCNC)
two-body channel $\tilde{t}_1\rightarrow c \chi$. The contribution to
$\tilde{t}_1\rightarrow  bf\bar f' \chi$
comes from the top quark exchange diagram and  the interference between top quark and sfermions
exchange diagrams because sparticles, except for $\tilde{t}_{1,2},\tilde{b}_1$ and $\chi$, are
decoupled in our scenario.
The flavor mixing of the lighter stop with charm-squark which can emerge from  radiative corrections
induces the lighter stop FCNC decay $\tilde{t}_1\rightarrow c \chi$. Their decay widths are given
by \cite{Hikasa:1987db,Das:2001kd,Hiller:2008wp,Boehm:1999tr}
\begin{eqnarray}
\Gamma(\tilde{t}_1\rightarrow  c\chi)&=&\frac{8}{9}\alpha|\epsilon |^2
\frac{\Delta m(\tilde{t}_1,\chi)^2}{m_{\tilde{t}_1}},\\
\Gamma_{\rm 4-body}\equiv\Gamma(\tilde{ t}_1\rightarrow bf \bar f' \chi)&=&
\mathcal{O}(10^{-5})\,\alpha^3\cos^2\theta_{\tilde{ t}}\,
\frac{\Delta m(\tilde{t}_1,\chi)^8}{m_{t}^2m_W^4m_{\tilde{t}_1}},
\end{eqnarray}
where $\epsilon$ is $\mathcal{O}(10^{-4})$ if all soft-breaking parameters have the same order of
magnitude.
It is clear that the four-body decay width increases more sharply with $\Delta m(\tilde{t}_1, \chi)$
than the FCNC two-body decay width, as shown in the right panel of Fig.~\ref{fig1}.
However, due to the ratio of $\Gamma_{\rm 4-body}/\Gamma(\tilde{t}_1\rightarrow  c\chi)$ is suppressed
by $\Delta m(\tilde{t}_1,\chi)^6/m_t^2m_W^4$ and the small coefficient, the four-body decay is not
competitive with the $\tilde{t}_1\rightarrow c \chi$ decay.

Because the soft $c$-jet from the $\tilde{t}_1\rightarrow c \chi$ decay is hard to detect, the  search
strategy for this coannihilation scenario is usually to exploit the $\tilde{t}_1\tilde{t}_1^*$
production in association with an energetic jet from the initial state radiation (ISR) which boosts
$\tilde{t}_1\tilde{t}_1^*$ system and produces large missing energy at the LHC.
The parton level events of the signal and backgrounds are generated with
\textsf{MadGraph5\_aMC@NLO} \cite{Madgraph}.
Then, the event parton showering and hadronization are performed by \textsf{Pythia} \cite{pythia}.
We use  \textsf{Delphes} \cite{delphes} to implement detector simulations where the anti-$k_t$ jet
algorithm and $\Delta R = 0.4 $ \cite{antikt} are set for the jet clustering.

To discriminate the signal and backgrounds, we require a leading jet with $p_T(j_1)>300$ GeV,
$|\eta|<2.4$ and azimuthal angle $\Delta\phi(j_1,\vec{p}_T^{\rm miss})>0.4$. We veto events with
electrons with $p_T>20$ GeV, $|\eta|<2.47$ or muons with $p_T>10$ GeV, $|\eta|<2.5$  to reduce
the $W(\rightarrow \ell \nu_\ell)j$ and $t\bar{t}$ backgrounds. Events having more than four jets
with $p_T>30$ and $|\eta|<2.8$ are vetoed.  The signal regions are defined with $E_T^{\rm miss}$ cuts:
300 GeV, 500 GeV, 700 GeV and 900 GeV.  The signal significance is calculated as $S/\sqrt{B}$ in
which the total background
$B=\sum\limits_i [B_i+(0.01B_i)^2]$ ($i=Z(\rightarrow\nu\bar{\nu})j$,
$W(\rightarrow \ell \nu_\ell)j$, $W(\rightarrow \tau \nu_\tau)j$),
where the systematic error on the backgrounds is set to $1\%$.

In the left panel of Fig.~1, we display the $2\sigma$ exclusion limits at the 13 TeV LHC
with ${\cal L}=36.1$ fb$^{-1}$ (the region on the left side of the curve is excluded)
 and the sensitivity at the 14 TeV LHC with ${\cal L}=3000$ fb$^{-1}$.
We can see that the current monojet search gives a loose limit
on the bino-LSP DM mass $m_{\chi}\simgt 260$ GeV and this limit can be raised to
570 GeV at the 14 TeV LHC with ${\cal L}=3000$ fb$^{-1}$.

\subsection{The $\tilde{t}_2 \tilde{t}^{\ast}_2$ production}
From the naturalness argument in Sec.~\ref{sec1}, $\tilde{t}_2$ can not be too heavy and
the $\tilde{t}_2\tilde{t}_2^*$ prodcution can be sizable at the LHC.
Since the LSP is bino-like in our scenario, the $\tilde{t}_2$ decay modes are mainly
$\tilde{t}_2\rightarrow \tilde{t}_1 Z$ and $\tilde{t}_2\rightarrow \tilde{t}_1 h$.
The corresponding decay widths are given by \cite{Bartl:1994bu}
\begin{eqnarray}
\Gamma(\tilde{t}_2\rightarrow  \tilde{t}_1Z)&\approx&
\frac{g_2^2\sin^22\theta_{\tilde{t}}m_{\tilde{ t}_2}^3}{256\pi m_W^2}\,
\lambda^{3/2}(m^2_{\tilde{ t}_2}, m^2_{\tilde{t}_1}, m^2_Z),\\
	\Gamma(\tilde{ t}_2\rightarrow \tilde{t}_1h)
&\approx&
\frac{g_2^2\cos^22\theta_{\tilde{ t}}m_t^2X_t^2}{64\pi m_W^2{m_{\tilde{t}_2}}}\; \lambda^{1/2}(m^2_{\tilde{t}_2}, m^2_{\tilde{t}_1}, m^2_h),
\end{eqnarray}
where $\lambda(a,b,c)=[1-(b+c)/a]^2-4bc/a^2$ is the kinematic factor.
In the limits $m^2_{\tilde{ t}_2}, m^2_{\tilde{t}_1}\gg m^2_{Z,h}$,
the factor $\lambda(m^2_{\tilde{ t}_2}, m^2_{\tilde{t}_1}, m^2_{Z,h})$
approximately equals to $(1-m^2_{\tilde{ t}_1}/m^2_{\tilde{ t}_2})^2$ and then
\begin{eqnarray}
\frac{\Gamma(\tilde{ t}_2\rightarrow \tilde{t}_1h)}{\Gamma(\tilde{t}_2
\rightarrow  \tilde{t}_1Z)}&\approx&\cos^2 2\theta_{\tilde{t}}=1
-\frac{4m^2_tX^2_t}{(m_{\tilde{t}_2}^2-m_{\tilde{t}_1}^2)^2}.
\end{eqnarray}

\begin{figure}[ht!]
	\centering
	\includegraphics[height=4in,width=4.5in]{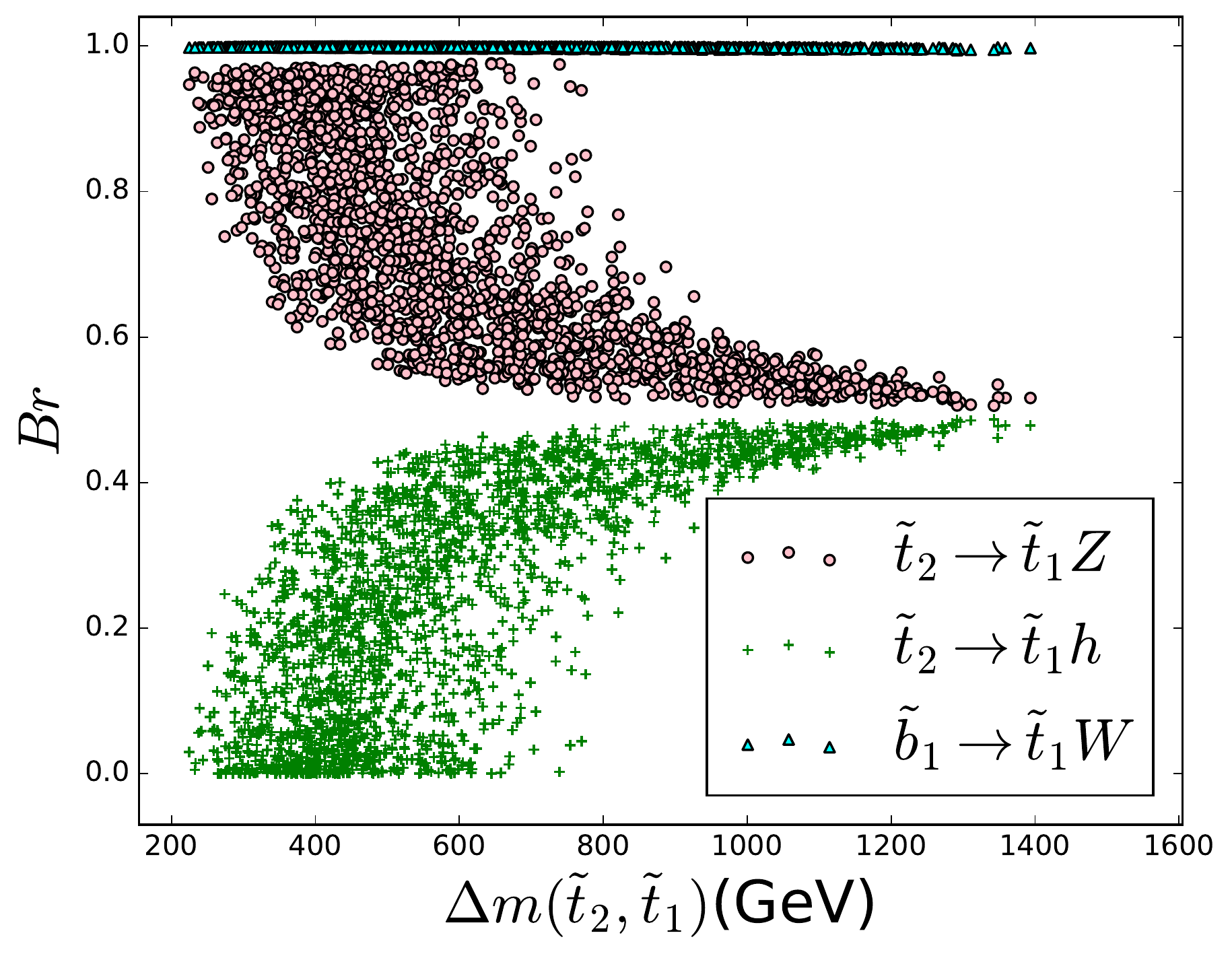}
	\vspace{-0.5cm}
	\caption{Same as Fig.\ref{fig1}, but showing the branching ratios
of $\tilde{t}_2\rightarrow  \tilde{t}_1Z$,
$\tilde{t}_2\rightarrow  \tilde{t}_1h$ and
$\tilde{b}_1\rightarrow  \tilde{t}_1W$.}
	\label{fig2}
\end{figure}
It should be noted that the decay width $\Gamma(\tilde{t}_2\rightarrow  \tilde{t}_1Z)$ is
always larger than $\Gamma(\tilde{t}_2\rightarrow  \tilde{t}_1h)$ even though the small loop
corrections are taken into account \cite{Bartl:1997pb,Bartl:1998xp}. In Fig.\ref{fig2},
we plot the branching ratio of $\tilde{t}_2\rightarrow  \tilde{t}_1Z$ and
$\tilde{t}_2\rightarrow  \tilde{t}_1h$. It is clear that
$\mathop{\rm Br}(\tilde{t}_2\rightarrow  \tilde{t}_1h)$ is lower
than $\mathop{\rm Br}(\tilde{t}_2\rightarrow  \tilde{t}_1Z)$
and their difference decreases with the mass splitting
$\Delta m(\tilde{t}_2,\tilde{t}_1)$ between heavier and lighter stops.
Since the masses of $\tilde{t}_1$ and the bino-LSP are nearly degenerate, the $\tilde{t}_1$
will appear as missing energy and the signal of $\tilde{t}_2\tilde{t}_2^*$ production at the LHC is
\begin{eqnarray}
pp\rightarrow \tilde{t}_2\tilde{t}_2^*\rightarrow ZZ+{E_T^{\rm miss}}
\quad\hbox{or}\quad Zh+E_T^{\rm miss}
\end{eqnarray}
where we neglect the $hh+E_T^{\rm miss}$ channel because its production rate is smaller than
the above channels. Here we investigate the $2\ell 2b$ final states, in which leptons come
from $Z$ decay and bottom quarks are from $Z/h$ decay, along with large missing energy.
The requirement of two leptons can efficiently reduce the QCD multi-jets
backgrounds\footnote{Tagging a soft $c$-jet from $\tilde{t}_1$ \cite{Ghosh:2013qga} or
boosted bosons $Z/h$ \cite{Kang:2017rfw}  may help to suppress the backgrouds.}.

\begin{figure}[ht!]
	\centering
	\includegraphics[height=3.3in,width=3.2in]{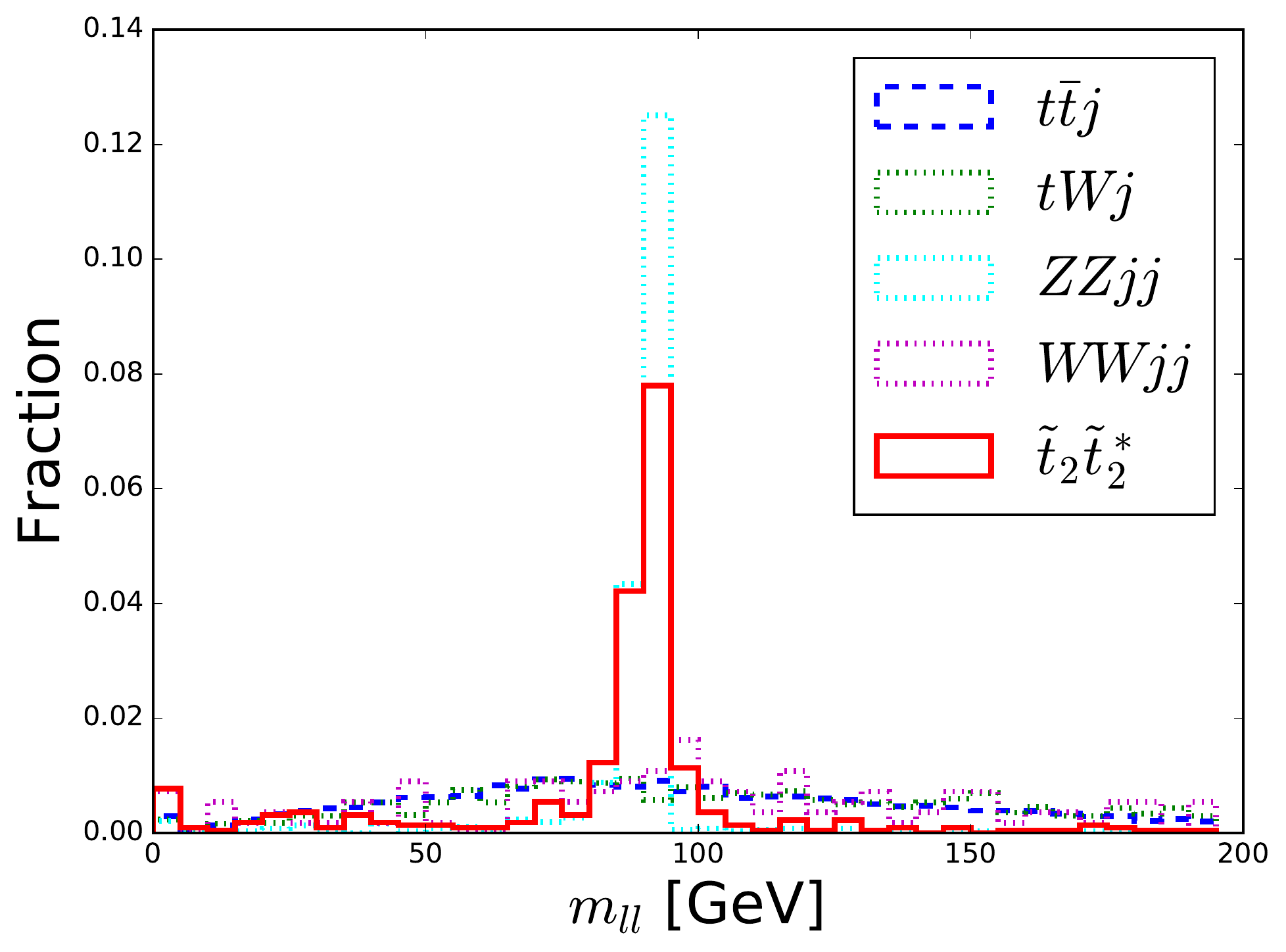}
	\includegraphics[height=3.3in,width=3.2in]{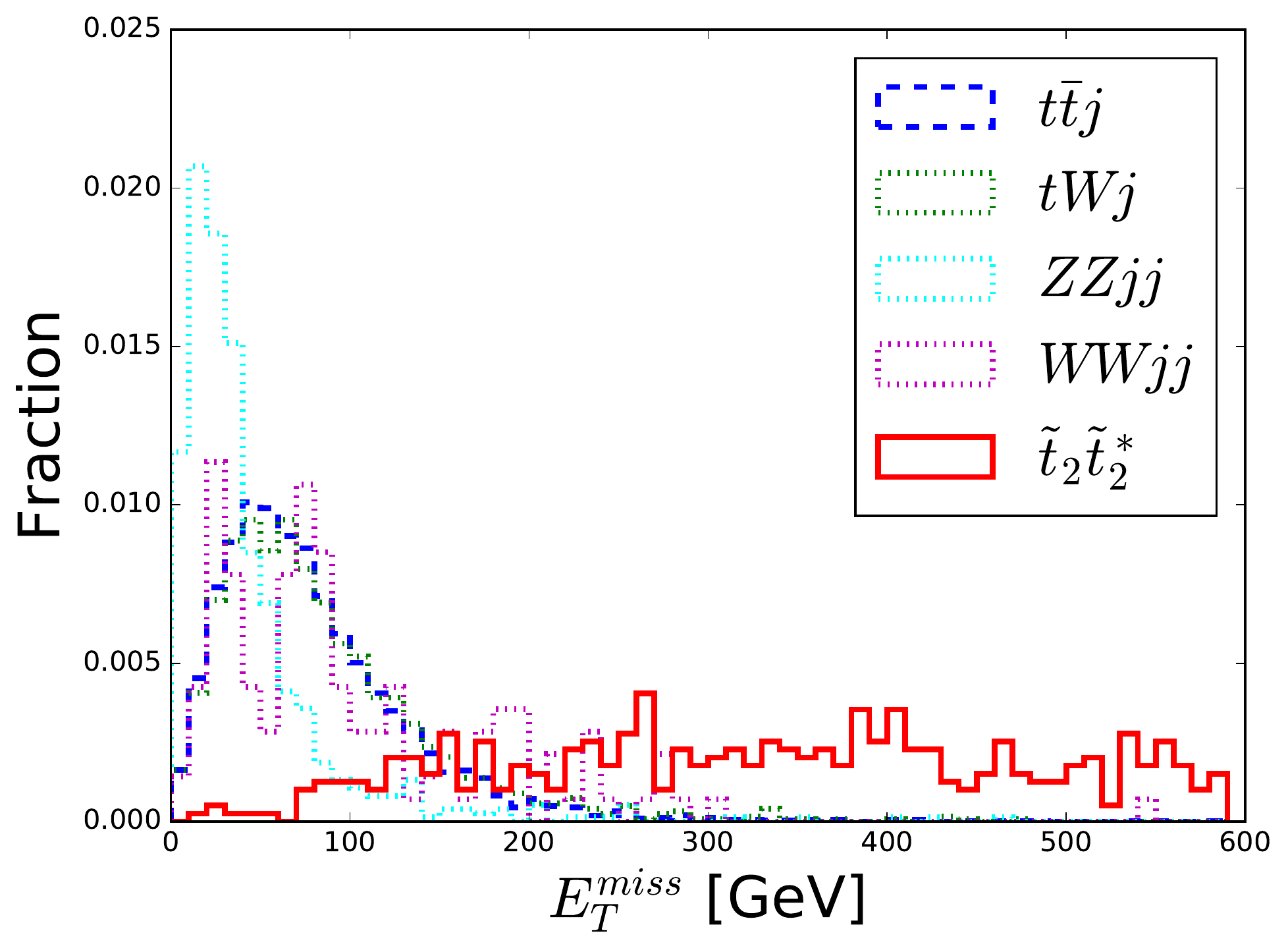}
	\vspace{-1.5cm}
\caption{The distributions of $m_{\ell\ell}$ and $E^{\rm miss}_T$ for backgrounds and signal at the 14 TeV LHC
after requiring exactly 2 leptons and 2 $b$-jets.  The signal benchmark point is chosen
as $(m_{\tilde{t}_2},m_{\tilde{t}_1},m_\chi)=(962,468,424)$ GeV.}
	\label{fig3}
\end{figure}

The main SM backgrounds are $t\bar{t}+{\rm jets}$, $tWj$, $ZZjj$ and $WWjj$. To discriminate the signal
and backgrounds, the following cuts are imposed:
\begin{itemize}
	\item[(i)] The event is required to have exact two leptons which form the opposite sign
and same flavor dilepton with $p_{T} (\ell) >30$ GeV and  $|\eta_\ell|<2.5$, where $\ell=e,\mu$.
According to the left panel of Fig.~\ref{fig3}, the invariant masses of dilepton should be
required in range of 80 GeV $<m_{\ell\ell}<$ 100 GeV to reconstruct $Z$ bosons.
	\item [(ii)]Jets must have $p_T(j) > 30$ GeV and $|\eta_j|<2.5$. We require  two $b$-jets and
the $b$-jet tagging efficiency is set to be $80\%$.
	\item[(iii)] From the right panel of Fig.~\ref{fig3}, the signal regions are designed
according to $ E_T^{\rm miss}$ cuts: 300~GeV, 350~GeV, 400~GeV, 450~GeV and 500 GeV.
\end{itemize}

\begin{table}[ht!]
\caption{The cut flow of events number for backgrounds and the signal at the HL-LHC.
The signal benchmark point is $(m_{\tilde{t}_2},m_{\tilde{t}_1},m_\chi)=(962,468,424)$ GeV. }
\begin{tabular}{|c|c|c|c|c|}
  \hline
  cut & \tabincell{c}{2\ leptons\\ $p^{\ell}_{T}>30$ GeV,$\left|\eta^{\ell}\right|<2.5$} & \tabincell{c}{2\ $b$-jets\\ $p^{b}_{T}>30$ GeV,$\left|\eta^{b}\right|<2.5$} & \tabincell{c}{$\left|m_{\ell\ell}-m_{Z}\right|<10$\\ $[$GeV$]$}&\tabincell{c}{$\slashed{E}_{T}>450$\\$[$GeV$]$}\\
  \hline\hline
   $t\bar{t}j$&1.248E+8&4.105E+7&5.849E+6&653\\
   \hline
   $tWj$&7.393E+6&1.349E+6&1.769E+5&30\\
  \hline
   $ZZjj$&7.212E+5&5.159E+4&4.531E+4&62\\
  \hline
   $WWjj$&2.603E+6&8.913E+4&1.506E+4&38\\
  \hline
   signal&3247&581&379&149\\
  \hline
  \end{tabular}
\label{table1}
\end{table}

\begin{figure}[ht!]
	\centering
	\includegraphics[height=4in,width=4.5in]{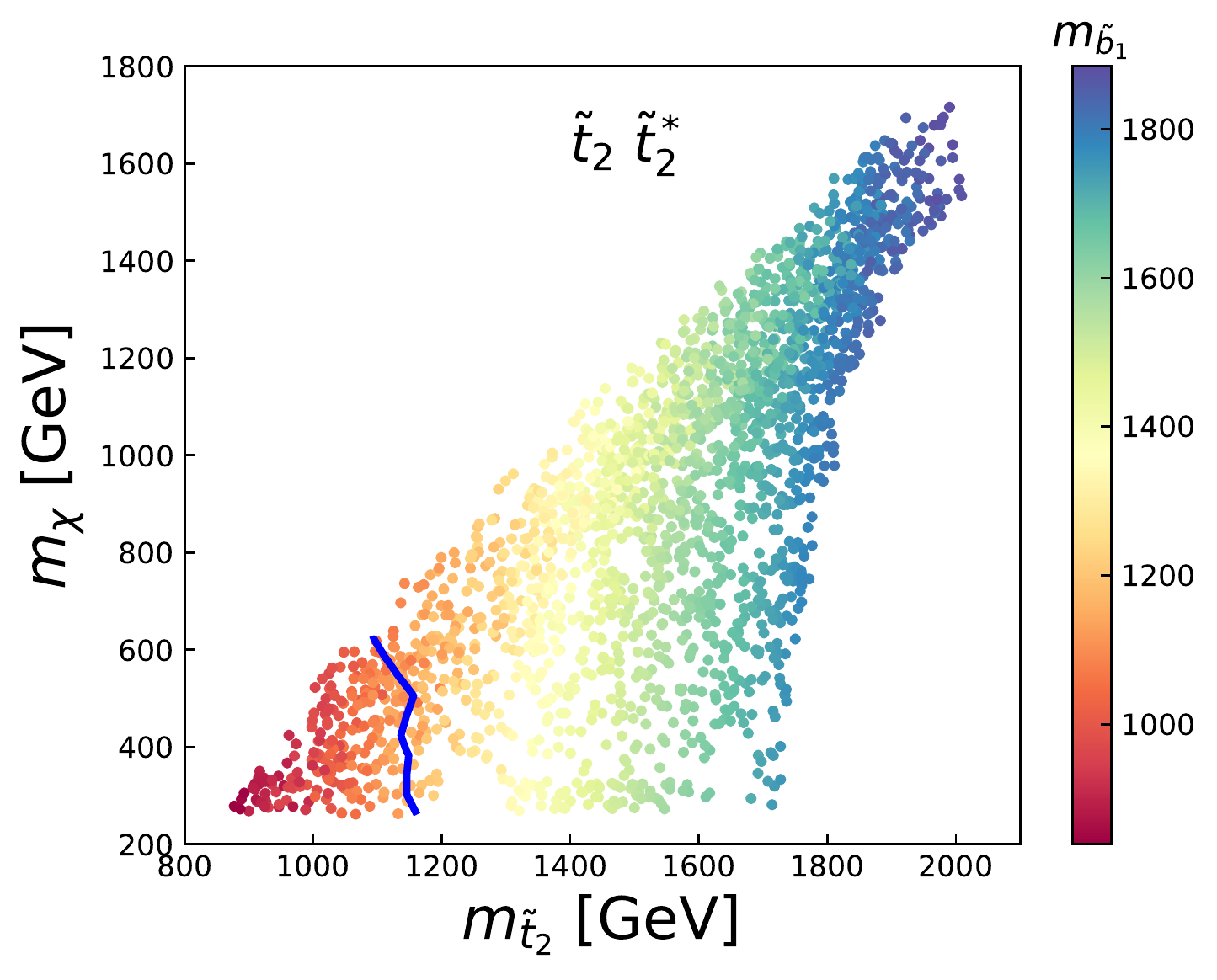}
	\vspace{-0.5cm}
\caption{Same as Fig.~1, but showing the observability of the $\tilde{t}_2 \tilde{t}^\ast_2$ production
on the ($m_{\tilde{t}_2}$,$m_\chi$) plane. The colormap represents the lighter sbottom mass $m_{\tilde{b}_1}$.
The blue curve is the $2\sigma$ significance and the left region has a sensitivity above $2\sigma$
level. }
	\label{stop2pair}
\end{figure}

In Table~\ref{table1}, a detailed cut flow of events number for backgrounds and the signal is
displayed. We can see that the $t\bar{t}j$ production is the largest SM background and the sum of
other backgrounds is also non-negligible. The requirement of the two-lepton invariant mass within
the range of 80--100 GeV can reduce the backgrounds by around $85\%$. It is clear that the cut
of $\slashed{E}_{T}>450$ GeV can remove backgrounds by near four orders of magnitude and this is
consistent with the distributions of the missing energy for backgrounds and the signal shown
in Fig.~\ref{fig3}. After imposing all these cuts, the significance
$S/\sqrt{B}$ for the benchmark point is about $5.32\sigma$.

In Fig.~\ref{stop2pair} we present the observability for the $\tilde{t}_2 \tilde{t}^\ast_2$ production.
The points to the left of the blue curve have a sensitivity above $2\sigma$ level and the colormap
shows the change in $m_{\tilde{b}_1}$. We can see that this stop pair production can cover
$m_{\chi}\simlt 600$ GeV for $m_{\tilde{t}_2}\simlt1100$ GeV at $2\sigma$ level.
This result is not sensitive to the mass splitting $\Delta m(\tilde{t}_1,\chi)$.

\subsection{The $\tilde{b}_1 \tilde{b}^{\ast}_1$ production}
The sbottom $\tilde{b}_1$  is lighter than the stop $\tilde{t}_2$ because of the mixing between
left and right handed stops. Since $\tilde{b}_1$ is left-handed in our scenario, it could decay
to the longitudal component of $W$ boson in association with $\tilde{t}_1$. The branching ratio
of  $\tilde{b}_1\rightarrow  \tilde{t}_1W$ is depicted in Fig.~\ref{fig2}. As we see,
$\tilde{b}_1$ dominantly decays to $W$ boson plus $\tilde{t}_1$. Then, the signal of
$\tilde{b}_1\tilde{b}_1^*$ production at the LHC is
\begin{eqnarray}
pp\rightarrow \tilde{b}_1\tilde{b}_1^*\rightarrow W^+W^- + E_T^{\rm miss}\;.
\end{eqnarray}

\begin{figure}[ht!]
	\centering
	\includegraphics[height=3.3in,width=3.2in]{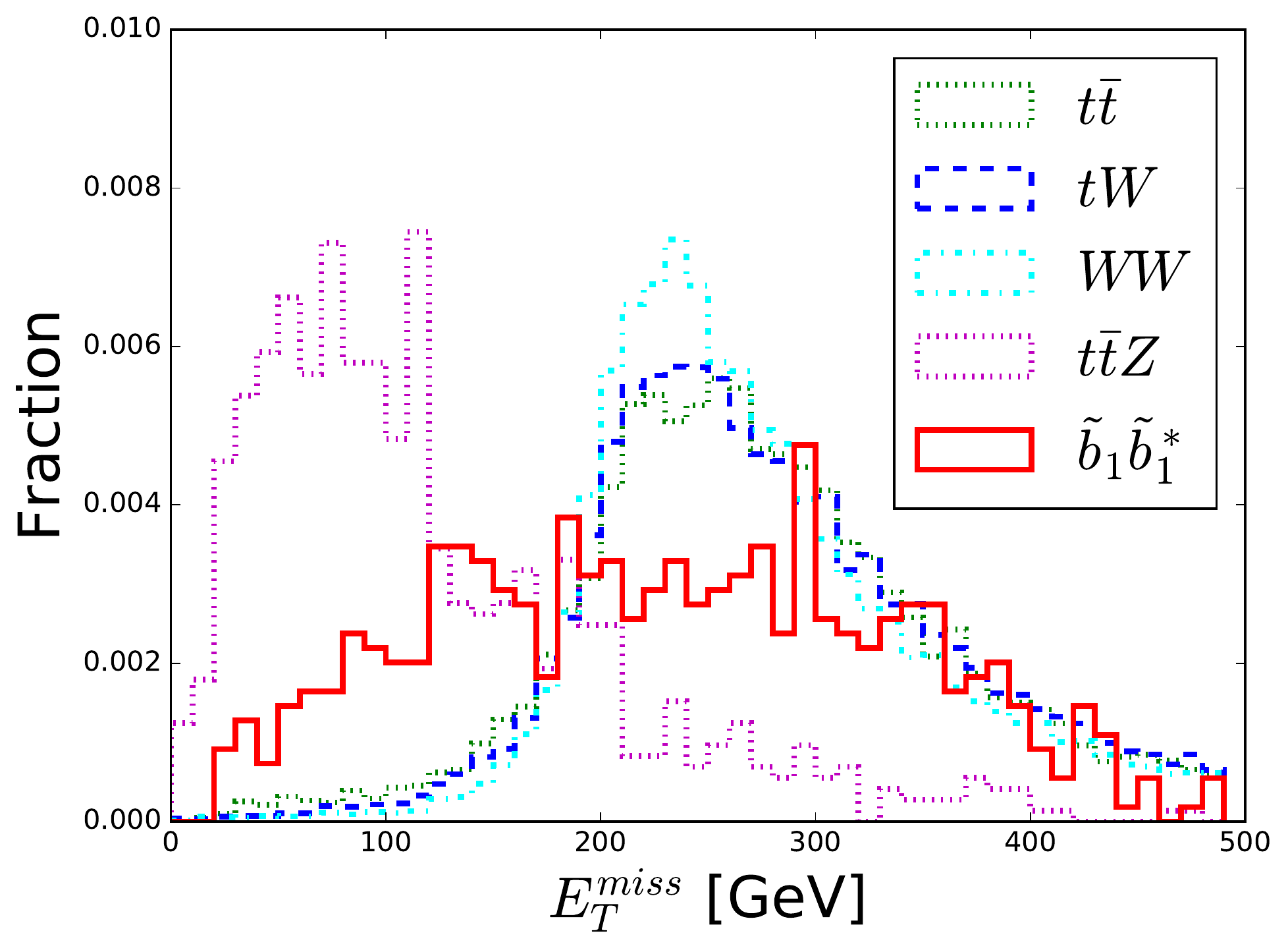}
	\includegraphics[height=3.3in,width=3.2in]{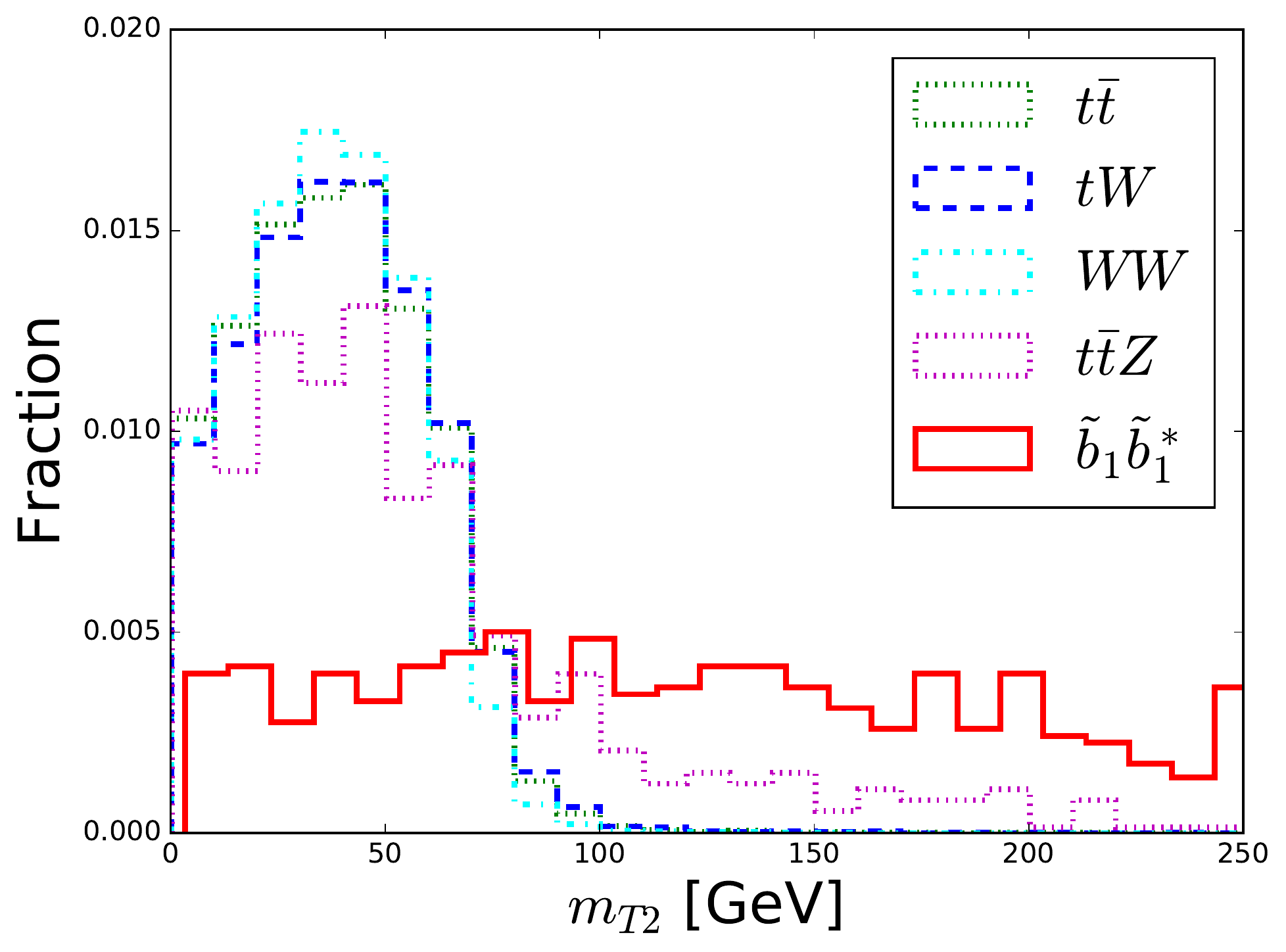}
	\vspace{-1.5cm}
\caption{The distributions of $E^{\rm miss}_T$ and $m_{T2}$ for backgrounds
and the signal at the 14 TeV LHC.  The signal benchmark point is
chosen as $(m_{\tilde{b}_1},m_{\tilde{t}_1},m_\chi)=(1200,910,870)$ GeV.}
	\label{figsbottomdis}
\end{figure}

\begin{table}[ht!]
\caption{The cut flow analysis of events number for backgrounds and the signal
at the HL-LHC. The signal benchmark point
is $(m_{\tilde{b}_1},m_{\tilde{t}_1},m_\chi)=(1200,910,870)$ GeV. }
\begin{tabular}{|c|c|c|c|c|c|}
  \hline
  cut &  $t\bar{t}$  & $tW$&$WW$&$t\bar{t}Z$  &signal\\
  \hline\hline
   $\sum\limits_\ell p_T^{\ell}>200$ GeV &251&82&234&2312&169\\
   \hline
  $p_T^b>50$GeV veto&119&54&226&810&142\\
  \hline
$ E_T^{\rm miss} >200$ GeV&112&52&221&179&103\\
  \hline
  \end{tabular}
\footnotesize
\label{tabsbottom}
\end{table}

\begin{figure}[ht!]
	\centering
	\includegraphics[height=4in,width=4.5in]{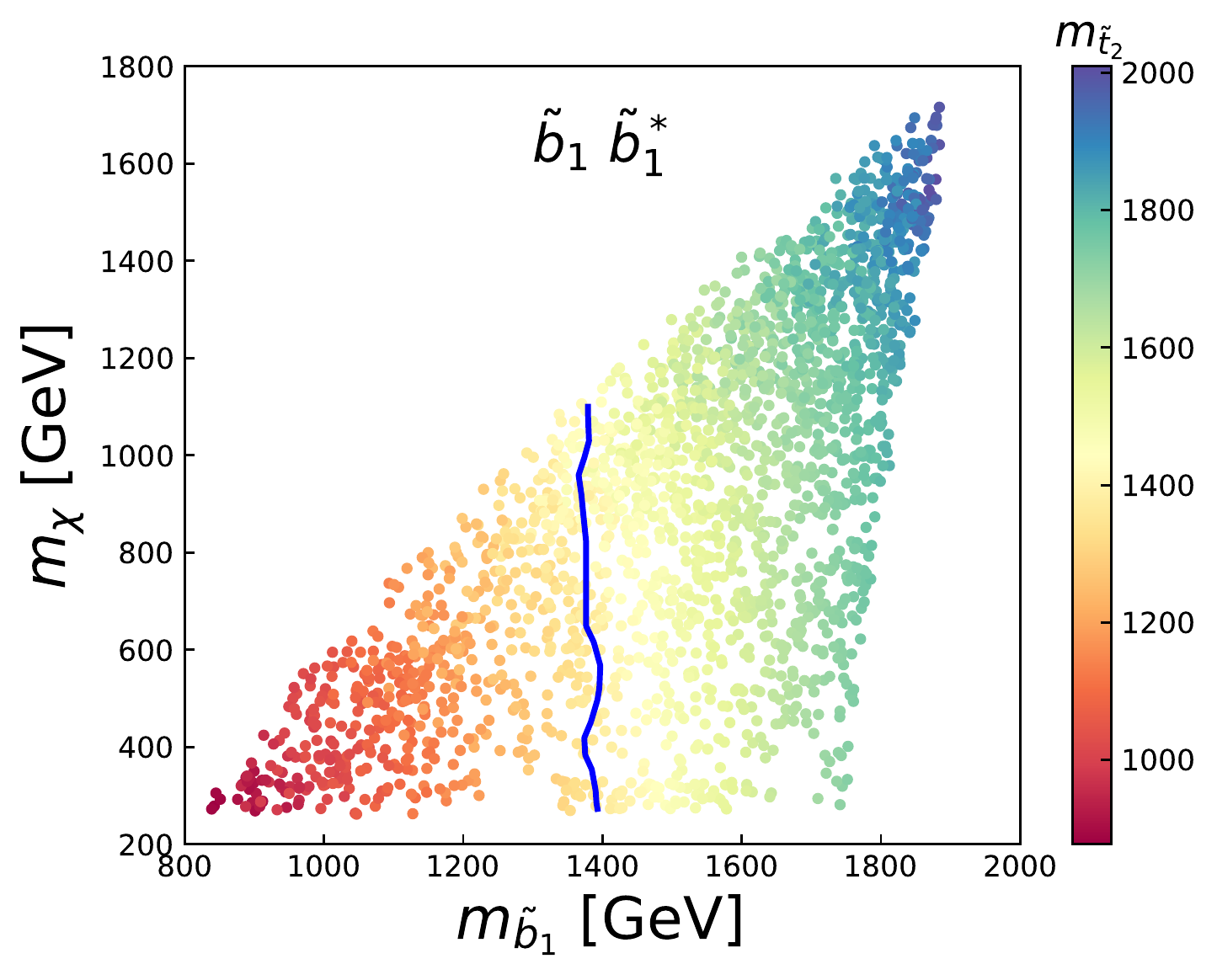}
	\vspace{-0.5cm}
\caption{Same as Fig.~1, but showing the observability of
the $\tilde{b}_1 \tilde{b}^\ast_1$ production on the ($m_{\tilde{b}_1}$,$m_\chi$) plane.
The colormap represents the change in $m_{\tilde{t}_2}$.
The blue curve is the $2\sigma$ significance and the left region has
a sensitivity above $2\sigma$ level.}
	\label{fig5}
\end{figure}
Similar to the search in case of $\mathop{\rm Br}(\tilde{b}_1\rightarrow  \tilde{t}_1W)=1$ \cite{An:2016nlb},
we investigate $2\ell E_T^{\rm miss}$  final state to probe this sbottom pair production.
We require exactly two opposite-sign leptons with $p_{T} (\ell) >25$ GeV and  $|\eta_\ell|<2.4$
to suppress the QCD multi-jet backgrounds. The invariant mass of dilepton is  required out
of the range $|m_{\ell\ell}-m_Z|<30$ GeV to remove $WZ$, $ZZ$ and $Z+$jets backgrounds.
Since the stop $\tilde{ t}_1$ boosts the $W$ boson, the sum of the two leptons' transverse momentum
$\sum\limits_\ell p_T^{\ell}>200$ GeV can be used for further seperating the signal from backgrounds.
Any $b$-jet with $p_T^b>50$ GeV is vetoed for suppressing $t\bar{t}$, $tW$ and $t\bar{t}Z$ backgrounds.
A detailed cut flow of events number for backgrounds and the signal is displayed in Table~\ref{tabsbottom}.
After the $\sum\limits_\ell p_T^{\ell}>200$ GeV requirement, we present the distributions of $E_T^{\rm miss}$
and  $m_{T2}$ of the two-lepton  system in Fig.~\ref{figsbottomdis}.  It is clear that
$E_T^{\rm miss}>200$ GeV can reduce the $t\bar{t}Z$ background efficiently. The variable $m_{T2}$
of the backgrounds has an endpoint around $M_W$. For a larger $m_{T2}$, the $t\bar{t}Z$ becomes
the dominant background. We seperate the signal regions according to $m_{T2}$ cuts: 100 GeV, 150 GeV
and 175 GeV.

We display the observability of the $\tilde{b}_1 \tilde{b}^\ast_1$ production in Fig.~\ref{fig5}.
It can been seen that such sbottom pair production can cover $m_{\chi}\simlt 1.1$ TeV
for $m_{\tilde{b}_1}\simlt1375$ GeV at $2\sigma$ level. Correspondingly, the lower bound
of $m_{\tilde{t}_2}$ can be pushed up to around 1.4 TeV. Therefore, this result is obviously
better than the $\tilde{t}_2 \tilde{t}^\ast_2$ production. This is mainly because this
sbottom pair production has 
a cleaner signal.

\section{CONCLUSIONS}
We have studied the stop-bino coannihilation region, in which the observed
dark matter relic abundance can be reproduced.  To test the scenario, we have
examined three correlated production processes
$\tilde{t}_1\tilde{t}_1^*+{\rm jet}$ (the $\tilde t_1$'s being invisible),
$\tilde{t}_2\tilde{t}_2^*$ and $\tilde{b}_1\tilde{b}_1^*$,
followed by the decays $\tilde{t}_2\to \tilde{t}_1+Z/h$ and $\tilde{b}_1\to \tilde{t}_1+W$,
at the HL-LHC{}.
Through Monte Carlo simulations for the signals and backgrounds, we found that
the $2\sigma$ sensitivity to the bino-like LSP can reach about 570 GeV, 600 GeV and 1.1 TeV
from the production process of $\tilde{t}_1\tilde{t}_1^*+ {\rm jet}$,
$\tilde{t}_2\tilde{t}_2^*$ and $\tilde{b}_1\tilde{b}_1^*$, respectively.
These three channels should be jointly considered at the future HL-LHC experiment.

\section*{ACKNOWLEDGMENTS}
We thank Lei Wu and Yang Zhang for helpful discussions.
This work was supported by the National Natural Science Foundation of China (NNSFC)
under grant Nos.11675242, 11821505, and 11851303, by Peng-Huan-Wu Theoretical
Physics Innovation Center (11847612), by the CAS Center for Excellence in Particle Physics
(CCEPP), by the CAS Key Research Program of Frontier Sciences and by a Key R\&D Program
of Ministry of Science and Technology under number 2017YFA0402204.

\end{document}